\documentclass[twoside]{articlek}

\usepackage[normal]{caption}

\renewcommand{\refname}{REFERENCES}

\usepackage{amsmath}

\def\BibTeX{{\rm B\kern-.05em{\sc i\kern-.025em b}\kern-.08em
            T\kern-.1667em\lower.7ex\hbox{E}\kern-.125emX}}

\usepackage{graphicx}

\begin{document}


\begin{flushleft}
{\large\bf 
Opportunities for nuclear reaction studies at future facilities}
\vspace*{25pt}

{\bf Martin Veselsky$^{1,}\footnote{E-mail: \texttt{martin.veselsky@savba.sk}}$,
Jozef Klimo$^{1}$, Nikoleta Vujisicova$^{2}$,  
and Georgios A. Souliotis$^{3,}\footnote{E-mail:
\texttt{soulioti@chem.uoa.gr}}$}\\
\vspace{5pt}
{$^1$Institute of Physics, Slovak Academy of Sciences, Dubravska cesta 9, 845 11
Bratislava, Slovakia\\
$^2$Faculty of Electronics and Informatics, Slovak Technical University, Bratislava, Slovakia\\
$^3$Laboratory of Physical Chemistry, Department of Chemistry,
National and Kapodistrian University of Athens, and
Hellenic Institute of Nuclear Physics,  Athens 15771, Greece}\\

\end{flushleft}


\vspace{5pt}
\begin{abstract}\noindent
Opportunities for investigations of nuclear reactions at
the future nuclear physics facilities 
such as radioactive ion beam facilities and high-power laser facilities are considered.
Post-accelerated  radioactive ion beams offer possibilities 
for study of the role of isospin asymmetry in the reaction mechanisms at various
beam energies. Fission barrier heights of neutron-deficient 
nuclei can be directly determined at low energies. Post-accelerated  radioactive ion beams, 
specifically at the future facilities such as HIE-ISOLDE, SPIRAL-2 or RAON-RISP 
can be also considered as a candidate for production 
of very neutron-rich nuclei via mechanism of multi-nucleon transfer. High-power 
laser facilities such as ELI-NP offer possibilities 
for nuclear reaction studies with beams of unprecedented properties. Specific
cases such as ternary reactions or even production 
of super-heavy elements are considered.    

\end{abstract}
\vspace{5pt}

\section{INTRODUCTION}

Progress in construction of advanced scientific infrastructure is 
a main driving force for progress in many fields of science, and 
this applies in particular to nuclear physics. Construction of still 
more powerful radioactive beam facilities allows to make spectacular 
progress in understanding of nuclear structure and 
post-acceleration of radioactive beams allows also to perform 
nuclear reaction studies using unstable beams. Besides radiactive 
beam facilities, high-power lasers emerge as another driving 
force of the progress in nuclear physics. While many concepts 
still need to be verified in order to perform detailed nuclear 
physics studies, use of high-power lasers may allow various 
types of experiments of interest for production of exotic nuclei 
and for nuclear astrophysics. In this proceeding we consider 
several possible experiments and extensions of capabilities 
at both radioactive beam and high-power laser facilities.

\section{(d,p)-transfer induced fission of heavy radioactive beams}

Nuclear fission was discovered 70 years ago and represents one of the most
dramatic examples of nuclear metamorphosis, whereby the nucleus splits into two
fragments releasing a large amount of energy. Fission is not only important for
applications such as the generation of energy and the production of
radio-isotopes, but also has direct consequences on the synthesis of the
heaviest elements in the astrophysical r-process, which is terminated by
fission, and on the abundance of medium-mass elements in the universe through
so-called "fission recycling" \cite{FissRecycl}. Furthermore, the fission process itself
enables the study of nuclear-structure effects in the heaviest nuclei. 
Until recently the low energy fission was studied in the region from thorium
to fermium using spontaneous fission, fission induced by neutrons and
light stable beams or using beta-delayed fission. Recently, the
probability of the electron-capture delayed  fission of 178,180Tl was measured
at ISOLDE and a new asymmetric mode of fission was observed \cite{FissMode}. 
One of the open questions in fission is the heigth of the fission barriers of
neutron-deficient nuclei. The region between lead and uranium is of special
interest since around the closed neutron shell N=126 the fission barrier height
is strongly influenced by shell structure, with direct implications to
predictions of production of super heavy nuclei, where fission barriers exist
purely due to shell structure. Statistical model calculations, used 
to reproduce experimental evaporation residue cross sections in this region
between lead and uranium, typically lead to extracted values of fission barrier
heigths in disagreement with theory, since available theoretical values \cite{Sierk,MollBF} 
need to be scaled down by 15 - 40 \%. The measured beta-delayed fission
probability was also used to deduce fission barrier height of the daughter
isotope 180Hg \cite{MVBF}, and deduced fission barriers were again 10 - 40 \% smaller than
theoretical estimates. However, since 180Hg (and other nuclei accessible in beta-delayed fission) 
is even-even nucleus, uncertainty remains concerning the magnitude of the pairing gap 
in saddle configuration and also concerning the extracted fission barrier height.

\begin{figure}[htbp]
\begin{center}
\includegraphics[width=7cm,clip]{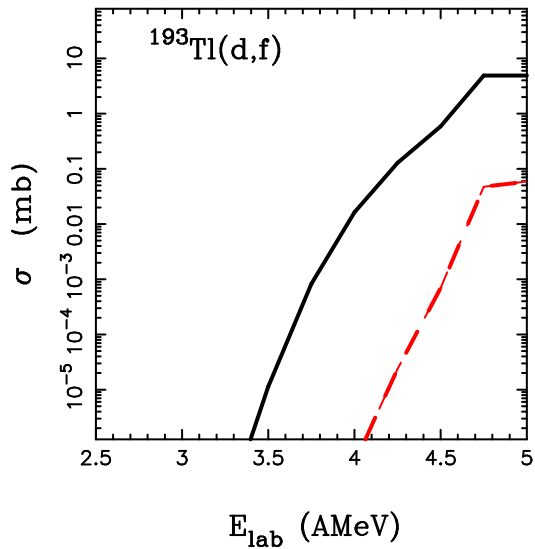}
\caption{
Fission cross sections for the radioactive beam 193Tl, calculated with
and without reduction of fission barrier \cite{Sierk} by 20 \% ( solid and dashed line,
respectively ). Strong sensitivity to fission barrier height offers possibility
to determine it experimentally.  
}
\label{fig193tl}
\end{center}
\end{figure}

The radioactive beams at the HIE-ISOLDE can be used to determine fission barrier
heights of exotic heavy fissile nuclei. Possibilities to observe fission
following the transfer reactions are investigated using the Talys code \cite{Talys}.  The
estimates in the region between lead and uranium show that energy upgrade of the
REX-ISOLDE post-accelerator to 4-5 AMeV will allow this type of low energy
fission studies. Specifically, it is of interest to observe transfer-induced
fission of odd elements such as Tl, Bi, At or Fr, since in this case the
estimated fission barriers will not be influenced by uncertainty in estimation
of the pairing gap in the saddle configuration, which is the case in
beta-delayed fission. Due to this
circumstance use of odd-Z beams is preferential, allowing to observe fission of
odd-odd nuclei, while use of even-Z beams may still allow to determine fission
barriers of even-odd nuclei, still more preferential than in beta-delayed
fission. It is possible to identify candidates for this type of measurement for
each of considered isotopic chains. Figure \ref{fig193tl} shows that fission cross sections
for the 193Tl radioactive beam increase dramatically when the fission barrier
height is scaled down by 20 \% (solid line) compared to standard values
of fission barriers \cite{Sierk} (dashed line). The isotope 193Tl appears especially
suitable to determine the fission barrier due to steep increase of the
excitation function and eventual availability of sufficient yield from the ISOL
target. In similar way, the beams of nuclei 199Bi, 201At and 209Fr can be
identified in analogous systematic estimates of fission cross sections of
corresponding odd elements. The observed fission rates of these beams can be
used to determine values of the fission barrier heights. 

Measurement at HIE-ISOLDE will be performed using the active
target ACTAR TPC \cite{ACTAR}, a time-projection chamber (TPC) filled with the target
(deuterium) gas. The use of ACTAR TPC offers several advantages, namely higher
observed fission rates and the possibility of obtaining the fission cross
sections for a range of beam energies in one measurement.
As an example, using deuterium gas with pressure 250 mbar, one obtains an
effective target thickness 1.6 mg/cm2 of deuterium. Assuming a target chamber
length parallel to beam axis of about 20 cm (corresponding to the  dimensions of
ACTAR TPC demonstrator), the beam would slow down from the initial energy of 5 AMeV to about
4.1 AMeV. This interval essentially covers the range of interest in the case of
transfer-induced fission of 193Tl. In ACTAR TPC the reaction vertex can be
reconstructed with a resolution better than 3 mm, allowing to measure more than
60 points of the excitation function over the energy range of interest. For
these points, a rate ranging from  about two events/minute at highest beam
energy down to one event per hour for the lowest energy can be calculated from
the amount of target material in the corresponding slice, the beam intensity
(10$^{6}$ pps) and a calculated cross section, with the fission barrier reduced by
20 \%. Integrated over the whole chamber, with average cross section around 2 mb
over the entire energy range, the fission rate can be estimated to twenty per
minute. Without reduction of the fission barrier, the expected fission rate can
be still estimated to some tens of fissions per hour. Thus the use of active
target ACTAR TPC provides the needed sensitivity, allowing to resolve the
long-standing question concerning the observed fission barriers of proton-rich
nuclei by way of their direct measurement. 
Observed fission rates will determine how detailed the investigations of the low
energy fission will be. Understandably, favorable will be lower values of
observed fission barrier heights, what appears quite probable for
neutron-deficient nuclei in the region around the shell closures Z=82 and N=126.
In such case, using the ACTAR TPC it will be possible to determine the mass
distribution of the fission fragments and thus asymmetry of the dominant fission
mode. The experiment was accepted as a part of the physics program of the HIE-ISOLDE 
and will be peformed during the year 2016. While the proposed experiment considers measurement of proton-rich nuclei, there is
also principal possibility to low-energy study fission the neutron-rich nuclei.
Good candidates for this type of study at the HIE-ISOLDE are the neutron-rich
radioactive beams such as 228Rn, of high interest for nuclear astrophysics
(study of r-process). 

\section{Production of exotic nuclei in peripheral nucleus-nucleus collisions
below 10 AMeV}

The fragmentation reactions offer a successful  approach to produce exotic nuclei
at beam energies above 100 AMeV, nevertheless they are restricted by the fact that
neutron excess is achieved by stripping the maximum possible number of protons
(and a minimum possible number of neutrons).
To reach an even higher neutron excess,  it is necessary to capture
additional neutrons from the target. Such an effect is observed in
reactions of nucleon exchange \cite{Volkov}  which dominate at beam energies
around the Fermi-energy (15--50 AMeV) \cite{MVNuclExch,MVProd,GSNrichEnh,GSKrSn,GS15AMeV}. 

In the Fermi-energy domain, peripheral nucleus-nucleus collisions are
described theoretically using
the mo\-del of deep-inelastic transfer, in combination with an appropriate model
of de-excitation.
Deep-inelastic transfer (DIT) occurs when the interaction of the projectile and the target
leads to formation of a di-nuclear configuration which exists long enough to allow
intense exchange of nucleons through a ``window" formed by the superposition of
the nuclear mean-fields in the neck region. Transfer of nucleons leads to gradual
dissipation of the kinetic energy of relative motion into internal degrees of freedom
such as intrinsic (thermal) excitation and/or angular momentum.
After re-separation, the hot projectile-like and
target-like primary fragments share approximately equal excitation energy
and undergo de-excitation via a cascade of
particle emissions or via simultaneous multifragmentation.

A very good description of experimental data from peripheral collisions
in the Fermi-energy domain was obtained \cite{MVNuclExch} using
the Monte Carlo deep-inelastic transfer (DIT) model of Tassan-Got \cite{TassanGot1,TassanGot2} 
for peripheral collisions,
combined with an appropriate choice of model description for central collisions. 
In the central collisions at Fermi energies, pre-equilibrium emission (PE) and
incomplete fusion (ICF) contribute to production
of the projectile-like fragments. 
The combined model framework is referred to as
the PE+DIT/ICF+SMM model \cite{MVProd}. 
Consistent good results can be obtained using the de-excitation 
code SMM \cite{SMM}, implementing the statistical model of multifragmentation (SMM) 
supplemented with particle evaporation and/or fission models
for the secondary emission stage.

In this context it is also of interest to investigate reactions
at energies around and below 10 AMeV and to establish to what extent the production rates
can be described using the model of nucleon exchange. 
Knowledge of the reaction mechanism at these low energies will allow us to
select the optimum projectile and target combinations, the appropriate target thickness,
as well as the optimum experimental setup 
for efficient production and collection of very neutron-rich exotic nuclei. 
The use of the thicker target, decelerating the beam particle to energies 
close to the Coulomb barrier, will, on one hand, further enhance the estimated intensities 
of secondary beams, as provided e.g. in refs. \cite{GSNrichEnh,GS-NIM03}, on the other 
hand it can simplify the set-up of the gas-cell necessary to stop the reaction products. 

\begin{figure}[htbp]
\begin{center}
\includegraphics[width=10cm,clip]{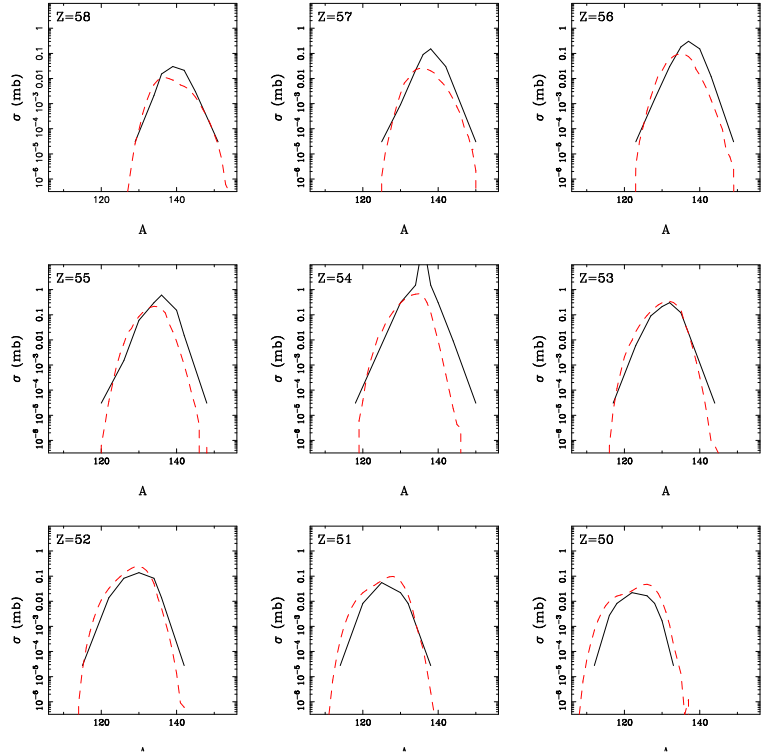}
\caption{
Simulations performed using model of deep-inelastic transfer, 
with modifications described in \cite{MV_LE} (dashed line) and 
compared to recent data in reaction of 136Xe+198Pt at 8 AMeV \cite{Watanabe} (solid line). 
}
\label{figxept8amev}
\end{center}
\end{figure}

The available experimental data from damped peripheral nucleus-nucleus collisions at beam energies
below 10 AMeV, specifically in reactions of 58,64Ni beams with Pb and U targets at beam energies around 6 AMeV \cite{NiPbLE,KrolasNiPb,NiUrLE} 
and in reactions of 22Ne beam with Zr, Th targets \cite{Artuch}, show that deep-inelastic transfer is the dominant reaction mechanism
leading to the production of projectile-like nuclei \cite{MV_LE}. As it was also demonstrated 
in the work \cite{MV_LE}, 
specific to this energy domain is a possible evolution of the extended nuclear
profile in the window (neck) region, primarily in reactions with very heavy target
nuclei. The effect seems to weaken with increasing beam energy, at 8 AMeV 
necessary extension of nuclear profile constitutes only 75 \% of the same 
at 6 AMeV and at 15 AMeV the effect disappears at all. 
Presence of this effect was further verified using the recently published data from 
reaction of 136Xe beam with 238U target at 8 AMeV \cite{Watanabe}. Again, using 
the same extension of nuclear profile as in \cite{MV_LE} for Ne + Th reaction at 8 AMeV, 
the experimental yields were reproduced rather well, as can be seen 
in Fig. \ref{figxept8amev} (due to presence of experimental 
yields rather than cross sectios a single normalization factor was employed). 
Thus the predictive power of the model simulations appears verfied and 
it can be used for predictions of achievable rates of very neutron-rich 
nuclei at the facilities with post-accelerated beams such as HIE-ISOLDE.

\begin{figure}[htbp]
\begin{center}
\includegraphics[width=10cm,clip]{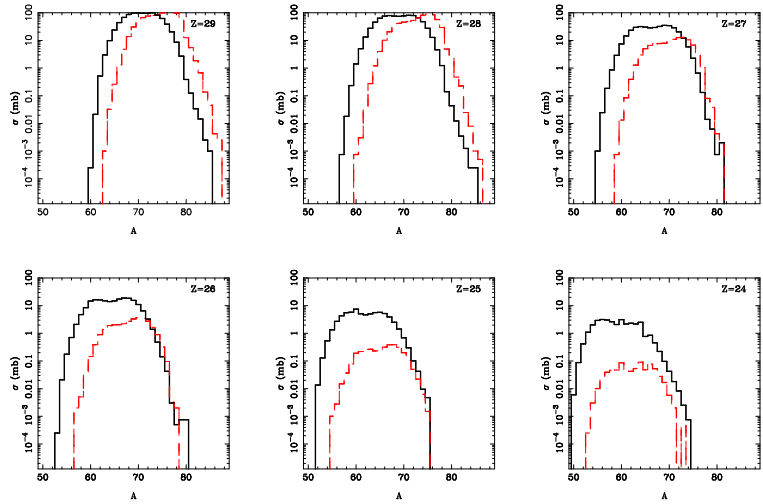}
\caption{
Simulations performed using model of deep-inelastic transfer,  
with modifications described in \cite{MV_LE} (dashed line)  
for reactions 74,78Zn+238U at 8 AMeV \cite{Watanabe} (solid and dashed line, respectively). 
}
\label{figznu8amev}
\end{center}
\end{figure}

As a first example, possibilities for production of the doubly magic nucleus 
78Ni can be investigated. Figure \ref{figznu8amev} shows results for the reaction 
of unstable nucleus 74Zn with uranium target at 8 AMeV (solid line), using the same extended 
profile as above. One can see that the production cross section for 
78Ni exceeds 1 mb, what assuming the presently achievable rate of low-energy 
RIB from primary spallation target of the order of 10$^{8}$/s, 
efficiency of post-acceleration process at the level of 5 \% and secondary 
target thickness of up to 10 mg/cm$^{2}$ leads to in-target production of 
one 78Ni nucleus in about five seconds. This rate can be further improved 
by upgrade to newly built proton linac and by further optimization of the 
yield from the spallation target. From the experimental point of view, 
a sensitive method allowing to collect all the products over wide 
angular range will be necessary, what favors the use of a gas-cell, 
where total efficiency of the order of 10 \% or more should be 
achievable. For the more neutron-rich RIB 78Zn (dashed line) the drop of the 
yields from the spallation target by about factor of 20 is practically compensated by 
increase of production cross section of 78Ni in secondary reaction and 
thus similar resulting yields of 78Ni can be expected.  

\begin{figure}[htbp]
\begin{center}
\includegraphics[width=10cm,clip]{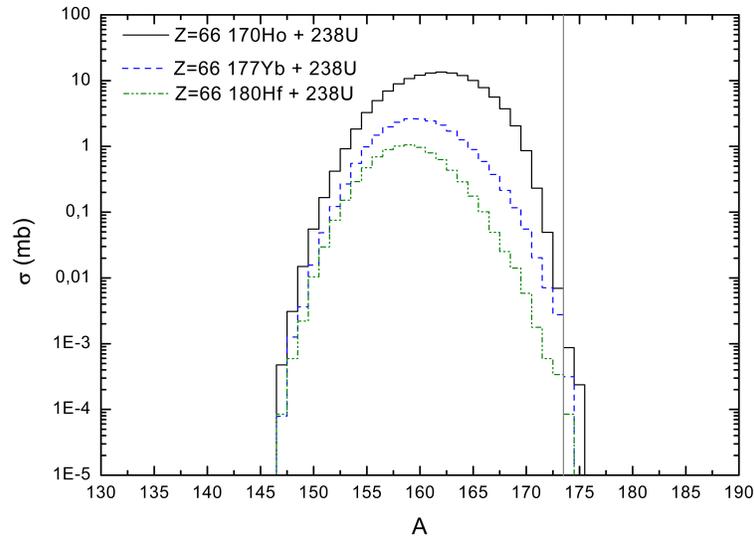}
\caption{
Lines show production cross sections of Dy isotopes from simulations performed using model of deep-inelastic transfer, 
with modifications described in \cite{MV_LE} (dashed line)  
for reactions 170Ho,177Yb,180Hf+238U at 8 AMeV \cite{Watanabe} (solid, dashed and dotted line, respectively). 
Vertical line marks heaviest known isotope. 
}
\label{fighyhu8amev}
\end{center}
\end{figure}

Besides the region of 78Ni, it is also interesting to know what can be 
achieved for heavier neutron-rich nuclei above heavier than fission 
fragments produced in fission of uranium. Figure \ref{fighyhu8amev} shows situation 
for production of isotopes of Dy in reactions of radioactive 
beams of 170Ho and 177Yb (with yields of low-energy beams of 10$^{8}$/s) 
with uranium at 8 AMeV (solid and dashed lines, respectively). The vertical line 
shows the heaviest known isotope. It appears, that many neutron-rich 
isotopes can be produced with reasonable rates and even the presently 
unknown isotopes appear reacheable. For comparison also stable beam 
180Hf is considered, where in principle higher primary beam rates 
can be considered, however it might require rotating target 
and also use of stable beam would result in much higher background 
from scattered beam in the gas-cell, from which the products of interest 
need to be separated. For even heavier nuclei, the secondary beam of 226Fr 
appears as good candidate for production of wide range of neutron-rich 
isotopes down to Z=80. 

Based on the above examples, it appears that the use of post-accelerated 
neutron-rich beams at HIE-ISOLDE for production of even more 
neutron-rich nuclei needs to be considered as an option for 
further upgrade.

\section{High-power laser as a tool for nuclear reaction studies} 

Basic reason for the implementation of high-power laser in nuclear 
physics is the eventual possibility to generate extremely high 
gradients of electric field, which can be used for acceleration 
of nuclei. As of now, electrons and nuclear particles with energy reaching several 
hundreds of MeV can be generated using the table-top laser with 
ultrashort pulses focused to energy densities 10$^{20}$ W/cm$^{2}$. 
Such kineitic energies are sufficient to initiate nuclear reactions and 
processes like photo-fission initiated by the laser were first 
observed at Rutherford Appleton Laboratory 
and Lawrence Livermore National Laboratory \cite{VULCAN,NOVA}, 
and more recently even using the table-top laser system \cite{Jena}. 
This field of nuclear physics thus can be considered as established. 

\begin{figure}[htbp]
\begin{center}
\includegraphics[width=14cm,clip]{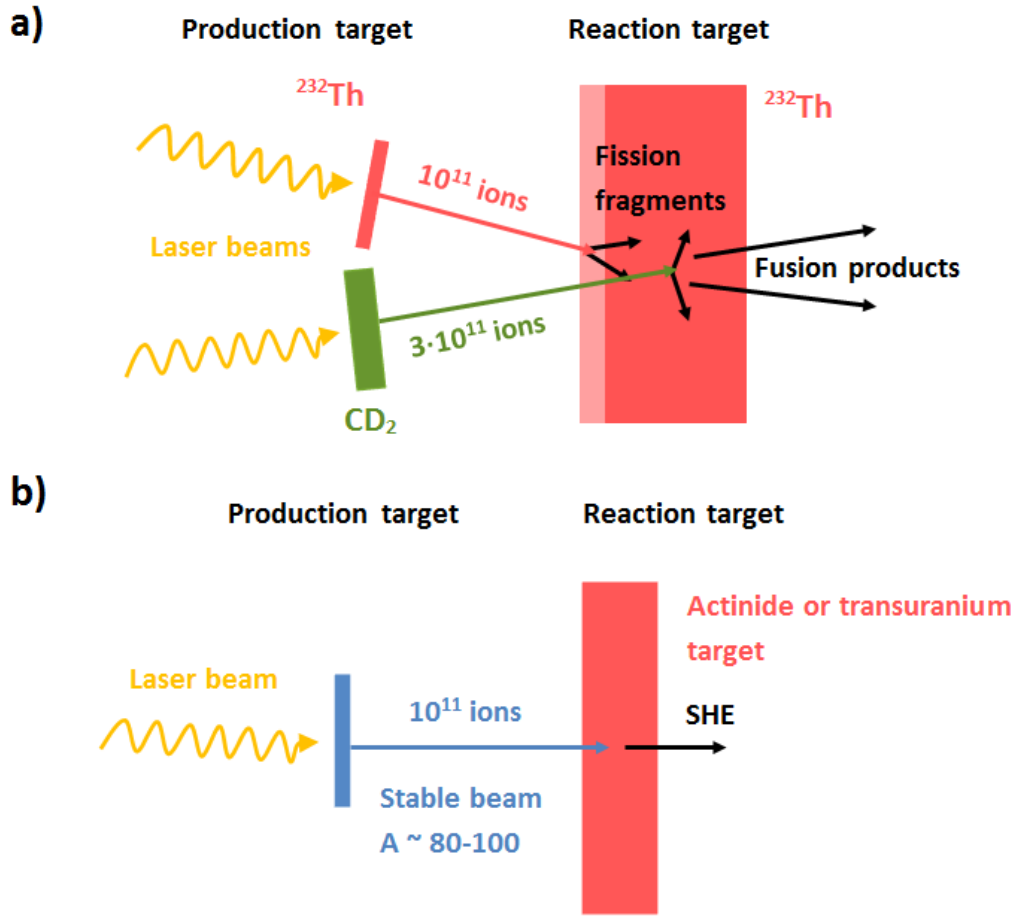}
\caption{
Scheme of the experiment (a) proposed by Habs et al. \cite{Habs} 
and (b) considered here for SHE production.
}
\label{figff}
\end{center}
\end{figure}

At the present, the most powerful laser is the BELLA Petawatt 
laser at Lawrence Berkeley National Laboratory, with the 
peak power of 1 PW (10$^{22}$ W/cm$^{2}$) and repetition rate of 1 Hz. 
At present time a major European research center ELI-NP is being constructed 
in Bucharest, as a nuclear physics branch of the 
ESFRI project ELI, focusing on the use of high-power lasers in material and 
nuclear physics. 
A laser system with peak power of 2 times 10 PW, 
worth 60 MEUR, will be installed in the nuclear physics branch. 
Among the proposed experiments there is an experiment proposed by the Habs et al. \cite{Habs},  
aiming to observe fusion of two unstable light fission fragments. 
Both fission fragments will be produced by intense 
laser pulses, one impinging on thorium foil, thus producing 
projectile-like fission fragments, and the other one impinging on CD2 foil, thus 
producing protons and deuterons, which will initiate fission 
of thorium in the target and thus production of target-like fission 
fragment. Scheme of this experiment is shown in Figure \ref{figff}a.

The proposal plans to use the so-called hole-boring variant of the 
Radiation Pressure Acceleration mechanism, allowing to accelerate 
the projectile-like light fission fragment to 7 AMeV, while 
the protons and deuteron are accelerated to the same energy 
per nucleon by another laser in order to initiate fission in Th target. 
Around 10$^{11}$ of Th projectile-like and 3 times more of light ions 
will be generated per laser burst. 
At an estimated purely geometric fusion cross section and 
for normal stopping in the target material, the proposal arrives to a rate  
of 1.5 fusion products per laser bunch, which is experimentally observable. 

The proposal further considers an effect of reduced stopping power 
for dense ion bunches in solid target and in this estimates production 
of more than 10$^{10}$ of projectile-like and target-like fission fragments 
per laser burst. At an estimated purely geometric fusion cross section 
the expected rate of fusion products rises to 4$\times$10$^{4}$ per laser burst.  
Due to uncertainty concerning the possible reduction 
of stopping the expected value is lowered to 10$^{3}$ per laser burst. 
The fusion products will be separated from the background 
using recoil separator and brought into its focal plane 
in order to identify them and study their properties. 
As a result of such experiment, properties of neutron-rich 
exotic nuclei in the vicinity of neutron shell N=126 can be explored, 
gaining valuable information for nuclear theory and nuclear astrophysics. 

Due to emerging limitations for production of further super-heavy elements 
using the contemporary accelerator technology such as cyclotrons and linacs, 
it is interesting to understand whether high-power laser can provide 
an alternative. Already in the above mentioned experiment, in principle 
also reactions of e.g. fission fragments with Th target nuclei can be 
imagined, however its observation will be disfavored by many orders of magnitude due to 
drop in the fusion cross sections. Also the possibility to produce 
heavy nuclei with Z=100 by fusion 
of two heavy fission fragments in a similar way as described 
above will be influenced by dramatic drop in cross sections. 

More conventional option would be to fuse the lighter stable beam 
with heavier Th-like (actinide or transurane) target nucleus.  
In the scenario with the stable beam-like nuclei and normal stopping 
(see Fig. \ref{figff}b) 
the thorium production foil will be replaced with lighter material 
and irradiated by the high-power laser. Similar number of accelerated 
nuclei should be obtained as in the case of thorium (thus gaining a 
factor of about $\times$10$^{3}$ when compared to beam-like light fission 
fragments) and the difference 
in yield of evaporation residues will be determined by the production 
cross section of SHE and by the fact that target-like nucleus does not need 
to fission (so another factor 4$\times$10$^{4}$ will be gained !). 
This means that e.g. for ER cross section of 1 pb  
the expected rate appears to be 4$\times$10$^{-5}$ per laser burst, 
with the expected repetition 
rate of 0.1 Hz (which is foreseen at ELI-NP) it will be one nucleus 
in about 100 hours ! Such rate would not be too far from the existing 
best facilities. 
However, it is known that excitation functions of production of super-heavy 
nuclei are quite narrow what may limit the expected rate. Typical thickness 
of heavy targets used for production of SHE is 0.5 mg/cm$^{2}$, corresponding 
to about 2000 layers of material. Main reason for such thin targets 
is the low velocity of evaporation residues and thus a relatively small 
range in the material. Use of thicker targets typically does not increase 
the observed rate of SHE and saturation is observed, thus demonstrating 
that only a limited thickness of about 0.5 mg/cm$^{2}$ contributes. 
As a result an additional factor of 10$^{-2}$ might be expected, 
thus reducing the expected rate in the case of normal stopping 
to 4$\times$10$^{-7}$ per laser burst or one nucleus in 10000 hours (more than one year) 
at production cross sections of 1pb. As mentioned above, 
for the production of SHE by the beam of light fission 
fragments (at normal stopping) the rate will be 10$^{3}$ times 
lower. 

In the case of reduced stopping the total Th target thickness is expected to grow by a factor of 100, 
and assuming that the "active" part of the target will grow also by factor of 100 (to 50 mg/cm$^{2}$)   
for stable beam particle one arrives to the above mentioned rate 4$\times$10$^{-5}$ one nucleus per 100 hours 
at production cross section of 1 pb, which is quite encouraging. For reduced stopping, the beam rate of the light fission 
fragment rate will be equal to the initial Th beam (and thus only by a factor of 3 lower than for stable beam) 
and one can expect one nucleus per 300 hours. Furthermore, one can expect an increase of fusion cross 
section for neutron-rich fission fragments which can make this option preferential to the use of 
stable beam. Of course these considerations depend on existence of the reduced stopping, 
so one can adopt more modest estimate by taking geometric mean of the options with 
and without reduced stopping. In that case the rates would be one nucleus per 1000 hours 
for stable beam particle and 55000 hours (more than 6 years) for light fission fragment. 

Thus one can conclude that the production of SHE using high-power laser 
is not excluded, however it strongly depends on the expected reduction 
of stopping in the dense ion bunches, which was not proved yet 
and for practical application also increase of laser power by 
several orders of magnitude would be necessary, in order to compensate 
also separation and detection efficiency. Still, the 
dynamical evolution of still more powerful lasers leaves much room 
for optimism. 

Besides other possible applications, it appears that high-power laser technique with 
ultra-dense beams might open pathway to ternary reactions of unstable 
nuclei e.g. ternary fusion of three light fission fragments. 
In the case of the above experiment, any of the 10$^{3}$ fusion 
products might fuse again and using similar considerations it appears 
that about 10$^{-4}$ of such ternary fusions can occurs per one laser 
bunch, about once per day at the ELI-NP setup. While it is still 
hardly observable, due to additional reduction of rate due to fission, 
laser facilities appear to provide environment to facilitate studies 
of ternary reaction in principle, due to high density of accelerated 
nuclei. This concept can be in principle tested e.g. using the 
accelerated Al nuclei hitting the Al target. If the high-density 
bunch of Al hits the target, fusion will occur and the fusion 
products will be able to fuse again. Assumming that the number of accelerated 
ions per laser bunch will be again of the order of 10$^{11}$ and taking into 
account that each atom of the target will be again Al, the probability 
of the ternary fusion will rise to the order of 10$^{3}$ per laser burst. 
This number was obtained assuming normal stopping, thickness of Al-target 
of 50 $\mu$m (as in the case of Th), and considering first two thirds 
of the Al-target as producing the fusion products which fuse again 
with Al-nucleus in the remaining third of the target. 
That should be obviously possible to observe.  
Of course the initial energy of the Al beam should be set so that 
the fusion products will be still fast enough to undergo fusion so 
some fine tuning of the above assumptions will be needed.  
Such experiment appears quite feasible and the ternary 
fusion may open pathway e.g. for simulation of astrophysical processes on Earth.

\section{CONCLUSIONS}

Future nuclear physics facilities 
such as radioactive ion beam facilities and high-power laser facilities 
offer many opportunities for investigations of nuclear reactions. 
Post-accelerated  radioactive ion beams offer possibilities 
for study of the role of isospin asymmetry in the reaction mechanisms at various
beam energies. Fission barrier heights of neutron-deficient 
nuclei can be directly determined at low energies. Post-accelerated  radioactive ion beams, 
specifically at the future facilities such as HIE-ISOLDE, SPIRAL-2 or RAON-RISP 
can be also considered as a viable candidate for production 
of very neutron-rich nuclei via mechanism of multi-nucleon transfer. High-power 
laser facilities such as ELI-NP, using the beams of unprecedented properties, 
offer possibilities for nuclear reaction studies such as ternary reactions and  
high power lasers with even higher intensities, which will be available in 
foreseeable future, can be even considered for production of super-heavy elements.

\noindent \\ACKNOWLEDGMENT:\\ 

This work is supported by the Slovak Scientific Grant Agency under contracts
2/0121/14, by the Slovak Research and Development Agency under contract
APVV-0177-11 (M.V.),   
by the NSFC of China under contract Nos. 11035009, 10979074, and 
by ELKE account No 70/4/11395 of the National and Kapodistrian University of Athens
(G.S.).

\end{document}